\documentclass[twocolumn]{aastex631}
\shorttitle{Hardening of Secondary CR Nuclei}
\shortauthors{Kawanaka \& Lee}
\graphicspath{{./}{figures/}}

\usepackage{newtxtext,newtxmath}
\usepackage[T1]{fontenc}
\usepackage{ae,aecompl}
\usepackage{hyperref}
\usepackage{graphicx}	

\begin{document}
\title{Origin of Spectral Hardening of Secondary Cosmic-Ray Nuclei}
\author[0000-0001-8181-7511]{Norita Kawanaka}
\affiliation{Department of Astronomy, Graduate School of Science, Kyoto University, \\
Kitashirakawa Oiwake-cho, Sakyo-ku, Kyoto, 606-8502, Japan}
\affiliation{Hakubi Center, Kyoto University, Yoshida-honmachi, Sakyo-ku, Kyoto, 606-8501, Japan}

\author[0000-0002-2899-4241]{Shiu-Hang Lee}
\affiliation{Department of Astronomy, Graduate School of Science, Kyoto University, \\
Kitashirakawa Oiwake-cho, Sakyo-ku, Kyoto, 606-8502, Japan}

\begin{abstract}
We discuss the acceleration and escape of secondary cosmic-ray (CR) nuclei, such as lithium, beryllium and boron, produced by spallation of primary CR nuclei like carbon, nitrogen, and oxygen accelerated at the shock in supernova remnants (SNRs) surrounded by the interstellar medium (ISM) or a circumstellar medium (CSM).  We take into account the energy-dependent escape of CR particles from the SNR shocks, which is supported by gamma-ray observations of SNRs, to calculate the spectra of primary and secondary CR nuclei running away into the ambient medium.  We find that if the SNR is surrounded by a CSM with a wind-like density distribution (i.e., $n_{\rm CSM}\propto r^{-2}$), the spectra of the escaping secondary nuclei are harder than those of the escaping primary nuclei, while if the SNR is surrounded by a uniform ISM, the spectra of the escaping secondaries are always softer than those of the escaping primaries. Using this result, we show that if there was a past supernova surrounded by a dense wind-like CSM ($\sim 2.5\times 10^{-3}M_{\odot}~{\rm yr}^{-1}$) which happened $\sim 1.6\times 10^5~{\rm yr}$ ago at a distance of $\sim 1.6~{\rm kpc}$, we can simultaneously reproduce the spectral hardening of primary and secondary CRs above $\sim 200~{\rm GV}$ that have recently been reported by AMS-02.
\end{abstract}


\section{Introduction}
Cosmic-ray (CR) secondary nuclei such as lithium, beryllium, and boron are considered to be produced via spallation of heavier nuclei such as carbon, nitrogen, and oxygen, which are mainly produced at Galactic supernova remnants (SNRs), during their propagation in the interstellar medium (ISM).  The amount of secondary CR nuclei produced per primary nucleus is proportional to the grammage traversed by the primaries in the ISM.  Therefore, the fluxes and rigidity dependence of secondary CRs are considered to be a probe of the propagation of CRs in our Galaxy.  The boron-to-carbon (B/C) ratio has been measured up to $\sim {\rm TeV}$ per nucleon by AMS-02 \citep{2016PhRvL.117w1102A}, and it has been shown that above the rigidity $R$ of $65~{\rm GV}$ the B/C ratio can be well fitted with a power law, $\propto R^{\Delta}$ with the index $\Delta \simeq -0.333$.  This agrees well with what is expected from the Kolmogorov turbulence theory, in which $\Delta$ is asymptotically equal to $-1/3$ \citep{1941DoSSR..30..301K}, and therefore it is suggested that the spatial diffusion coefficient $D$ as a function of rigidity is proportional to $R^{1/3}$.  However, AMS-02 later reported that the spectra of CR lithium, beryllium, and boron deviate from a single power law above $\rm 200~{\rm GV}$ in an identical way and that they harden even more than the primary CRs \citep{PhysRevLett.120.021101} which have already been known to harden at higher rigidities from measurements by PAMELA \citep{2011Sci...332...69A}, CREAM \citep{2010ApJ...714L..89A}, and AMS-02 \citep{2015PhRvL.114q1103A,2015PhRvL.115u1101A,PhysRevLett.119.251101}.

Several scenarios have been discussed to account for the deviation of the index of the secondary-to-primary ratio at high rigidities from what is expected from the diffusion coefficient in the bulk of the ISM, including the effect of propagation in the Galaxy \citep{2010PhRvD..82b3009C,2012ApJ...752L..13T,2014ApJ...786..124C,2016ApJ...827..119C,2015PhRvD..92h1301T}, re-acceleration of secondary CRs \citep{2019MNRAS.488.2068B, 2020JCAP...11..027Y}, contribution from different kinds of sources \citep{2018PhRvL.120d1103K, 2020ApJ...889..167B, 2021ChPhC..45d1004N, 2021ApJ...911..151M} and the effect of acceleration of secondary CRs in supernova remnants (SNRs) \citep{2009PhRvL.103h1104M,2009PhRvD..80l3017A,2013PhRvD..87d7301K,2014PhRvD..90f1301M,2014PhRvD..89d3013C}.  These models are proposed mainly to explain the positron fraction measured by PAMELA and AMS-02 \citep{2009Natur.458..607A, 2013PhRvL.110n1102A}, the antiproton-to-proton ratio measured by AMS-02 \citep{2016PhRvL.117i1103A}, or the spectral hardening of proton and helium \citep{2011Sci...332...69A, 2015PhRvL.114q1103A, 2015PhRvL.115u1101A}. Especially, in the last scenario, secondary CR nuclei are produced inside the SNRs via the spallation of primary CR nuclei being accelerated at the SNR shock, and are accelerated at the same shock.  Since the spectra of secondary CR nuclei produced in this way are harder than that of primary CR nuclei, if their contributions are comparable to the background CR fluxes above a certain rigidity, the rigidity dependence of the secondary-to-primary ratio would deviate from what is expected from simple ISM diffusion.  Very recently, \cite{2020arXiv201212853M} performed a comprehensive study on this scenario by searching for the best fit parameters to account for the recent AMS-02 measurements of CR protons, helium, positrons, anti-protons, and primary/secondary nuclei.  However, they discuss the spectra of secondary CRs that are shock-accelerated and advected to the downstream (i.e., inside the SNRs), and they do not consider the spectra of CRs escaping from the SNRs, which are what we observe.

In this paper we discuss the energy-dependent escape of primary and secondary CRs that are accelerated in the SNR shock, and predict their energy spectrum assuming that they come from a local SNR.  The escape of CR particles from an SNR has been intensively investigated to interpret the observed CR spectra and the gamma-ray emission feature of SNRs interacting with molecular clouds \citep{2005A&A...429..755P,2009ApJ...694..951R,2009MNRAS.396.1629G,2010MNRAS.407.1773C,2010A&A...513A..17O,2011MNRAS.410.1577O,2012MNRAS.427...91O,2016PhRvD..93h3001O,2011ApJ...729L..13O,2011MNRAS.415.1807D,2011ApJ...729...93K,2013MNRAS.431..415B}.  If CRs escape the SNR in an energy-dependent way, the spectrum of escaping CRs would be steeper than that of CRs trapped inside the SNR \citep{2010A&A...513A..17O,2010MNRAS.407.1773C}, which may account for the steepness of the observed CR spectral index.  Therefore, it is worth investigating theoretically the production, acceleration, and escape of secondary CRs in an SNR, and comparing them with observational data.  In this work, we consider CR accelaration in SNRs not only within a uniform ISM but also those surrounded by a dense circumstellar medium (CSM).  Recent intense optical and near-infrared transient searches have revealed that many SNe show signs of strong interaction between their ejecta and the CSM surrounding them.  There are various classes of CSM-interacting SNe: Type IIn, Ibn, Ia-CSM, type I superluminous SN, etc (for a comprehensive review, see \citealp{2017hsn..book..403S}).  The origin of such dense CSM is still uncertain, but many observations indicate that their progenitors had expelled the stellar material shortly before their explosions, and the mass-loss rates estimated from their observational properties are up to $\sim 10^{-3}M_{\odot}~{\rm yr}^{-1}$ (assuming their wind velocities are $\sim 100~{\rm km}~{\rm s}^{-1}$), continuing for decades \citep{2014MNRAS.439.2917M}.  The non-thermal emissions of photons or neutrinos due to the strong interactions between an SN ejecta and a dense CSM have been investigated in some studies \citep{2016APh....78...28Z, 2019ApJ...874...80M, 2019ApJ...872..157W, 2019ApJ...885...41M, 2020MNRAS.494.2760C, Matsuoka_2020}, and the nature of such a system as the origin of CRs has also been investigated \citep{2011PhRvD..84d3003M, 2014MNRAS.440.2528M}.  However, the production and acceleration of secondary CRs in such a system, as well as their escape into the ISM, have not been discussed yet.

In Section 2, we describe our model of CR production inside a SNR surrounded by a CSM and energy-dependent escape of primary and secondary CRs.  In Section 3 we solve the transport equations for primary and secondary CR nuclei to evaluate the spectra of the escaping CRs, compare them with observations, and discuss our results in light of stellar evolution scenarios.  We conclude our work in Section 4.

\section{Model}
\subsection{Energy-dependent Cosmic-ray Escape Scenario}
Here we describe how CR particles accelerated at the SNR shock escape the source into the ISM depending on their energy, following \cite{2010A&A...513A..17O}.  In the context of the diffusive shock acceleration (DSA) theory \citep{1987PhR...154....1B}, particles can go back and forth across the shock front and gain kinetic energy because they are scattered by the turbulent magnetic field around the shock.  Especially, the turbulence in the upstream (i.e. the outside of the SNR shock) may be amplified by the streaming instability caused by the accelerating particles themselves \citep{1978MNRAS.182..147B,2000MNRAS.314...65L}.  Since the turbulence is generated only in the vicinity of the shock front, if accelerated particles have sufficiently high energy, they can escape the SNR into the far upstream region without being trapped by the turbulent magnetic field.  Using the diffusion coefficient of accelerated particles as a function of their momentum $D(p)$, the shock velocity $u_{\rm sh}$, and the distance from the shock beyond which the turbulence is negligible $l$, the escape condition for a particle accelerated at the shock can be described as
\begin{eqnarray}
\frac{D(p)}{u_{\rm sh}}\gtrsim l, \label{escapecondition}
\end{eqnarray}
where the left hand side represents the diffusion length of a particle with momentum $p$.  \cite{2008ApJ...678..939Z} investigated the generation of magnetohydrodynamic (MHD) turbulence driven by the non-resonant streaming instability using numerical MHD calculations, and show that $l$ should be the same order as the radius of the SNR, $R_{\rm sh}$.

As a SNR evolves with time, the magnetic field around the shock decays and the speed of the SNR shell slows down.  As a result, the escape condition Eq.(\ref{escapecondition}) also evolves with time.  Especially, if the diffusion length of a particle with a specific momentum increases faster than the SNR radius ($\sim l$), the required momentum of particles for escaping the SNR decreases with time.  In other words, CR particles accelerated at the SNR shock can escape the SNR to far-upstream in an energy-dependent way.  In the following subsections, we will evaluate the energy distribution of CR particles produced in a SNR taking into account the spallation of primary CRs, the production of secondary CRs, and their energy-dependent escape.  

\subsection{Primary and Secondary Cosmic-rays in the Supernova Remnant Shock}
Here we present our formalism to derive the energy distributions of primary and secondary CRs produced in the SNR and their escaping fluxes as functions of time.  In the following discussion, we focus on the CR nuclei of lithium, beryllium, boron, carbon, nitrogen, and oxygen.  Among them, only oxygen is regarded as a pure primary element, while other lighter elements are partly or entirely produced via spallation of heavier nuclei during their propagation.  Such an approximation is often used in calculating the fluxes of light nuclei \citep{2009PhRvL.103h1104M,2021ChPhC..45d1004N}.

We assume that the CR particles can be regarded as test particles during DSA in a SNR.  Letting the shock front be at $x=0$, the stationary transport equation for the distribution functions of CR nuclei $f_i(x,p)$ ($i$ represents the type of nuclei) in the shock rest frame is
\begin{eqnarray}
u(x)\frac{\partial f_i}{\partial x} &=& \frac{\partial}{\partial x}\left[ D_i(p)\frac{\partial f_i}{\partial x} \right] + \frac{p}{3}\frac{du}{dx}\frac{\partial f_i}{\partial p}-\Gamma_i f_i + q_i \nonumber \\
&&+ u_-Q_i\delta(x)\delta(p-p_0), \label{transport}
\end{eqnarray}
where $u(x)$ is the fluid velocity, $D_i(p)$ is the diffusion coefficient for a nuclei of $i$ with momentum $p$, $\Gamma_i$ is the total spallation rate of a nuclei $i$ (i.e. $\Gamma_i=\sum_{i > j}\Gamma_{i\rightarrow j}$), $q_i$ is the source term due to the spallation of parent particles, and $Q_i$ is the injection rate of a nuclei $i$ at the shock front (the injection momentum is $p_0$).  Considering that the kinetic energy per nucleon of a nucleus is conserved before and after the spallation, $q_i$ is given by
\begin{eqnarray}
4\pi p^2 \frac{dp}{d\varepsilon_k} q_i(p)=\sum_{i<j}\Gamma_{j\rightarrow i}N_j(\varepsilon_k),
\end{eqnarray}
where $N_j(\varepsilon_k)=4\pi p^2f_j(p)(dp_j/d\varepsilon_k)$ ($p_j$ is the momentum of a nucleus $j$ with kinetic energy of $\varepsilon_k$) is the kinetic energy distribution function of nuclei $j$, and $\Gamma_{j\rightarrow i}$ is the rate at which a nucleus $i$ is produced via spallation of a heavier nucleus $j$.  Here we adopt the table of the spallation cross sections in \cite{1990acr..book.....B}.  The fluid velocity is given by
\begin{eqnarray}
u(x) = \begin{cases}
u_- & (x<0) \\
u_+ & (x>0),
\end{cases}
\end{eqnarray}
where $u_-=u_{\rm sh}$ and $u_+=u_{\rm sh}/r$ are constants, and $r$ is the shock compression ratio, which is assumed to be equal to 4 (the strong shock limit, ignoring non-linear effects from CR feedback) hereafter.

We then solve the transport equation (\ref{transport}) by imposing the following boundary conditions:
\begin{eqnarray}
&{\rm (i)}& \lim_{x\rightarrow -0}f_i = \lim_{x\rightarrow +0}f_i, \\
&{\rm (ii)}& \lim_{x\rightarrow -l} f_i = 0, \\
&{\rm (iii)}& \left| \lim_{x\rightarrow +\infty} f_i \right| < \infty, \\
&{\rm (iv)}& \left[ D_i(p)\frac{\partial f_i}{\partial x} \right]^{x=-0}_{x=+0} \nonumber \\
&=&\frac{1}{3}(u_+-u_-)p\frac{\partial f_{i,0}}{\partial p}+u_-Q_i \delta(p-p_0), \label{bc4}
\end{eqnarray}
where condition (i) means that the distribution functions should be continuous across the shock, and condition (ii) means the free escape of CR particles from the outer boundary. Condition (iv) comes from the integration of Eq.(\ref{transport}) across the shock front ($x=0$), and it yields the differential equation for the distribution function at $x=0$ with respect to $p$, $f_{i,0}(p)$.  Hereafter we solve Eq.(\ref{transport}) for the relativistic regime, i.e., the kinetic energy per nucleon $\varepsilon_k$ is greater than a few ${\rm GeV}/{\rm n}$, so that we can approximate $p\approx A\varepsilon_k/c$.  

Following \cite{2009PhRvL.103h1104M} we can solve for the energy distribution function of nuclei $i$, $N_i(\varepsilon_k)=4\pi p_i^2 f_i(p)(dp_i/d\varepsilon_k)$ separately in the downstream ($x>0$) and the upstream ($x<0$), where we can neglect the second and fifth terms on the right hand side of Eq.(\ref{transport}).  In the downstream and upstream, the solutions of Eq.(\ref{transport}) are described respectively as
\begin{eqnarray}
N_i^+&=&\sum_{j\ge i} E_{ji}e^{\lambda_j x/2}, \\
N_i^-&=&\sum_{j\ge i} F_{ji}e^{\kappa_j x/2} +G_i,
\end{eqnarray}
where
\begin{eqnarray}
\lambda_j&=&\frac{u_+}{D_j}\left( 1-\sqrt{1+4D_j\Gamma_j/u_+^2} \right), \\
\kappa_j&=&\frac{u_-}{D_j}\left( 1+\sqrt{1+4D_j\Gamma_j/u_-^2} \right),
\end{eqnarray}
and $E_{ji}$ and $F_{ji}$ are determined recursively as
\begin{eqnarray}
E_{ji}&=&\frac{-4\sum_{m\le j}\Gamma_{m\rightarrow i}E_{jm}}{D_i\lambda_j^2-2u_+\lambda_j-4\Gamma_i}  ~ ({\rm for}~j>i), \\
E_{ii}&=&N_{i,0}-\sum_{j>i}E_{ji}, \\
F_{ji}&=&\frac{-4\sum_{m_\le j}\Gamma_{m\rightarrow i}F_{jm}}{D_i\kappa_j^2-2u_-\kappa_j-4\Gamma_i} ~ ({\rm for}~j>i),\\
F_{ii}&=&\frac{N_{i,0}-\sum_{j>i}F_{ji}(1-e^{-\kappa_j l/2})}{1-e^{-\kappa_i l/2}}, \\
G_i&=&N_{i,0}\left( 1-\frac{1}{1-e^{-\kappa_i l/2}} \right)-\sum_{j> i}F_{ji}\frac{e^{-\kappa_j l/2}-e^{-\kappa_i l/2}}{1-e^{-\kappa_i l/2}},
\end{eqnarray}
where $N_{i,0}=4\pi p_i^2f_{i,0}(p_i)(dp_i/d\varepsilon_k)$.  The differential equation for the distribution function at the shock front $f_{i,0}$ can be derived from the boundary condition (iv) as
\begin{eqnarray}
p\frac{\partial f_{i,0}}{\partial p}
&=&-\frac{3D_i}{2(u_- -u_+)}\left[ \left( \frac{\kappa_i}{1-e^{-\kappa_i l/2}}-\lambda_i \right)f_{i,0} \right. \nonumber \\
&& \left. +\sum_{j>i} \left\{ \tilde{F}_{ji} \left( \kappa_j-\frac{1-e^{-\kappa_j l/2}}{1-e^{-\kappa_i l/2}}\kappa_i \right) - \tilde{E}_{ji} \left( \lambda_j-\lambda_i\right) \right\} \right] \nonumber \\
&& +\frac{3u_-}{u_- -u_+}Q_i \delta(p-p_0),
\end{eqnarray}
where
\begin{eqnarray}
\tilde{E}_{ji}&=&\frac{E_{ji}}{4\pi p_i^2 (dp_i/d\varepsilon_k)}, \\
\tilde{F}_{ji}&=&\frac{F_{ji}}{4\pi p_i^2 (dp_i/d\varepsilon_k)}.
\end{eqnarray}
The difference between \cite{2009PhRvL.103h1104M} and this study is the position of the (effective) escape boundary: the former assumes that $f_i$ should damp at $x=-\infty$, while we impose the outer boundary condition $f_i=0$ at a finite distance, $x=-l$.  When the acceleration timescale is much shorter than the spallation timescale (i.e., $\Gamma_i D_i/u_-^2 \ll 1$) and the escape boundary is very far from the shock front (i.e., $e^{-\kappa_i l/2} \ll 1$), this equation is asymptotically identical to Eq. (17) in \cite{2009PhRvL.103h1104M}.

We can evaluate the escape flux of CR nuclei from the SNR at $x=-l$ as
\begin{eqnarray}
\phi_i(p)&=&u_- f_i|_{x=-l}-D_i(p)\left. \frac{\partial f_i}{\partial x}\right|_{x=-l} \nonumber \\
&=&-D_i(p)\sum_{j\ge i}\frac{\kappa_j}{2} \tilde{F}_{ji}e^{-\kappa_j l/2},
\end{eqnarray}
which is the function of time through the size of the escape boundary, $l$.  The energy spectrum of CR particles escaping the SNR per unit time is given by
\begin{eqnarray}
\frac{dN_{{\rm esc},i}(\varepsilon_k)}{dt}\simeq 4\pi R_{\rm sh}^2 \cdot 4\pi \left( \frac{A\varepsilon_k}{c} \right)^2\left| \phi_i (p)\right| \frac{dp_i}{d\varepsilon_k}, \label{escaping}
\end{eqnarray}
where $R_{\rm sh}$ is the radius of the SNR shock.

To understand the nature of the escape flux $\phi_i(p)$, let us assume that the $i$-nuclei are purely primary (i.e., they are not produced via spallation of heavier nuclei), and that the loss due to spallation is negligible.  In this case, the escape flux can be described as
\begin{eqnarray}
-\phi_i(p)=\frac{u_- f_{i,0}(p)}{\exp (u_- l/D_i)-1},
\end{eqnarray}
Here we assume Bohm-type diffusion, in which the mean free path of a charged particle is proportional to its Larmor radius, inside the SNR:
\begin{eqnarray}
D_i(p)=\eta_g\frac{c^2p}{3ZeB},
\end{eqnarray}
where $B$ is the magnetic field strength and $\eta_g$ is the gyro factor.  We can see that the momentum at which the absolute value of the escape flux of the $i$-nuclei attains its maximum value is given by
\begin{eqnarray}
p=p_{i,m}\equiv \frac{3u_- l Z e B}{\gamma c^2 \eta_g}, \label{pmax}
\end{eqnarray}
and the maximum value of the escape flux is 
\begin{eqnarray}
-\phi_i(p_{i,m})=\frac{u_- f_{i,0}(p_{i,m})}{e^{\gamma}-1},
\end{eqnarray}
\citep{2009MNRAS.396.2065C, 2010A&A...513A..17O}.  Noting that the distribution function at the shock front behaves as $f_{i_0} \propto p^{-\gamma}$ when $p\ll p_{i,m}$ where $\gamma=3u_-/(u_- - u_+)=3r/(r-1)$, and that $f_{i,0}(p)\propto \exp(-p/p_{i,m})$ when $p\gg p_{i,m}$, the particles accelerated at the SNR shock with momentum $p_{i,m}$ can escape the SNR most efficiently, i.e., $p_{i,m}$ can be regarded as the maximum momentum of the escaping particles.

We can also consider the particles that are accelerated at the shock but cannot escape the shock are advected downstream.  These particles are confined in the SNR and are finally released into the ISM as CRs when the SNR is disrupted.  The energy spectrum of this CR component can be evaluated in the following way.  First, the number of the advected particles per unit momentum per unit time is given by 
\begin{eqnarray}
\frac{dN_{\rm adv}(p)}{dt}=4\pi R_{\rm sh}^2\cdot 4\pi p^2 f_{i,0}(p)\frac{u_{\rm sh}}{r},
\end{eqnarray}
where $u_{\rm sh}/r$ is the fluid velocity in the downstream in the shock rest frame.  The momentum loss of particles due to the expansion of the SNR (i.e., adiabatic cooling; \citealp{2005A&A...429..755P}) is described as $dp/dt= -(u_{\rm sh}/R_{\rm sh})p$.  Assuming the shell radius expands as $R_{\rm sh}\propto t^q$ where $q$ is a constant, the momentum evolves with time as $p \propto t^{-q}$.  As a result, the energy spectrum of CR particles that were once advected downstream and are released at the end of the SNR's life is given by
\begin{eqnarray}
&&\int_{t_{\rm ini}}^{t_{\rm fin}} dt^{\prime} \frac{dN_{{\rm adv},i}(\varepsilon_k)}{dt^{\prime}}  \nonumber \\
&\simeq& \int_{t_{\rm ini}}^{t_{\rm fin}} dt^{\prime} 4\pi R_{\rm sh}^2 \cdot 4\pi \left(\frac{A\varepsilon_k}{c}\right)^2 f_{i,0}(p^{\prime})\frac{u_{\rm sh}}{r}\frac{p^{\prime 2}dp^{\prime}}{p^2dp},
\end{eqnarray}
where $p^{\prime}$ is the momentum at the time of injection $t^{\prime}$ of a particle whose momentum is $p$ at the time of release into the ISM, $t_{\rm fin}$ (i.e., $p^{\prime}=p(t^{\prime}/t_{\rm fin})^{-q}$).

To evaluate the resulting CR spectra, we need a model for the spatial diffusion coefficient, $D_i(p)$, which in general depends on time because of the evolution of the magnetic field around the shock, and the primary CR injection rate at the shock, $Q_i$.  In the next section, we describe a phenomenological model to determine these quantities based on analytic formulae of SNR evolution and the observed spectra of Galactic CRs.

\subsection{Evolution of SNRs and CR fluxes}

Neglecting the back reaction of particle acceleration, the dynamics of the SNR shock can be well described by analytic solutions.  When an SNR is surrounded by the ISM or a CSM, the SNR expands freely until it sweeps up a mass comparable to its own mass from the surrounding medium, and afterward the SNR shell is decelerated and expands in a self-similar way.  This is so called the Sedov-Taylor phase, and the evolution of the SNR shell radius is determined by the explosion energy of a supernova $E_{\rm SN}$, the ejecta mass $M_{\rm ej}$, and the number density of the ambient medium $n$.  In the case with the uniform ISM, the SNR shell radius is given by
\begin{eqnarray}
R_{\rm sh}=R_{\rm S} \left( \frac{t}{t_{\rm S}} \right)^{2/5}
\end{eqnarray}
where $R_{\rm S}$ and $t_{\rm S}$ are the Sedov radius and Sedov time, respectively, and they are given by
\begin{eqnarray}
R_{\rm S}&=&4.59~{\rm pc}\left( \frac{M_{\rm ej}}{1M_{\odot}} \right)^{1/3} \left( \frac{n}{0.1~{\rm cm}^{-3}} \right)^{-1/3}, \\
t_{\rm S}&=&450~{\rm yr}\left(\frac{E_{\rm SN}}{10^{51}~{\rm erg}} \right)^{-1/2}\left( \frac{M_{\rm ej}}{1M_{\odot}} \right)^{5/6} \left( \frac{n}{0.1~{\rm cm}^{-3}} \right)^{-1/3},
\end{eqnarray}
while in the case that the ambient density profile is wind-like (i.e. $n\propto r^{-2}$), the SNR shell radius is given by
\begin{eqnarray}
R_{\rm sh}= R_{\rm S} \left( \frac{t}{t_{\rm S}} \right)^{2/3}
\end{eqnarray}
where
\begin{eqnarray}
R_{\rm S}&=&1.26~{\rm pc}\left(\frac{M_{\rm ej}}{1M_{\odot}} \right)\left( \frac{\dot{M}}{10^{-3}M_{\odot}~{\rm yr}^{-1}} \right)^{-1}\left( \frac{v_w}{100~{\rm km}~{\rm s}^{-1}} \right), \\
t_{\rm S}&=&178~{\rm yr}\left(\frac{E_{\rm SN}}{10^{51}~{\rm erg}} \right)^{-1/2}\left( \frac{M_{\rm ej}}{1M_{\odot}}\right)^{3/2}\left( \frac{\dot{M}}{10^{-3}M_{\odot}~{\rm yr}^{-1}} \right)^{-1} \nonumber \\
&&\times \left( \frac{v_w}{100~{\rm km}~{\rm s}^{-1}} \right),
\end{eqnarray}
where $\dot{M}$ and $v_w$ are the mass loss rate and wind velocity, respectively.  Here we use the density profile for a steady wind, i.e., $n=\dot{M}/(4\pi m_p v_w r^2)$.

We here adopt a phenomenological model proposed by \cite{2009MNRAS.396.1629G} (see also \citealp{2010A&A...513A..17O}), which is based on the assumption that Galactic SNRs are responsible for CRs with energy below the knee ($\sim 10^{15.5}~{\rm eV}$).  In this model it is assumed that the maximum momentum of a CR particle $p_{i,m}$ accelerated at the SNR is limited by its escape during the Sedov-Taylor phase, and that the maximum energy of an escaping particle, $\sim cp_{i,m}$, at the beginning of the Sedov phase is equal to the knee energy ($\sim 10^{15.5}~{\rm eV}$).  Using a variable $\chi$ to describe the evolution of an SNR (e.g., the SNR's age, the shock radius of an SNR, etc.), we assume that the maximum momentum of an escaping CR particle decreases with time as $p_{i,m}\propto \chi^{-\alpha}$, and that the spectrum of CRs accelerated at the SNR is proportional to $\chi^{\beta}p^{-s}$ (i.e., CRs inside the SNR have a power-law spectrum with index of $s$ and their total number increases with time).  One can then see that the spectrum of CRs escaping the SNR is proportional to $p^{-s+\beta/\alpha}$.  This means that if $\alpha$ and $\beta$ are positive (i.e., the maximum momentum decreases and the total number of accelerated CRs increases with time), the spectrum of escaping CRs become steeper than that inside the SNR.  Here $\alpha$ and $\beta$ are the phenomenological parameters that are determined so that the resulting CR spectra are consistent with observations.  Following \cite{2010A&A...513A..17O}, we hereafter adopt $R_{\rm sh}$ as $\chi$, and fix the parameters as $\alpha=6.5$ and $\beta=2.1$.  Actually, with these parameters, the maximum energy of escaping CRs evolves from the knee energy at the beginning of the self-similar phase to $1~{\rm GeV}$ at the end of the Sedov phase, when the radius of the SNR shell becomes 10 times larger than $R_S$, and the spectral index of escaping primary CRs would be similar to what is inferred from observations, assuming that the diffusion coefficient in the ISM, $D_{\rm ISM}(R)$, is proportional to a power of rigidity, $R$.

One can determine the diffusion coefficient and the normalization factor of the escaping CR flux in the following way.  First, the time dependence of $p_{i,m}$ ($\propto \chi^{-\alpha}$) is related to the evolution of the diffusion coefficient, which depends on the amplification and decay of magnetic field and the turbulence around the shock \citep{2003A&A...403....1P, 2005A&A...429..755P, 2012ApJ...745..140Y}.  Assuming the Bohm diffusion inside the SNR, one can determine the evolution of the diffusion coefficient in the SNR using the relation in Eq.(\ref{pmax}).  Assuming $l\propto R_{\rm sh}$, we can describe the evolution of the diffusion coefficient at a specific momentum as $\propto \chi^{\alpha-1/2}$ in the case with uniform ISM, while $\propto \chi^{\alpha+1/2}$ in the case with the wind-like ambient density profile. Second, the normalization factor of the spectrum of CRs accelerated at the SNR ($\propto \chi^{\beta}$) is related to the CR injection rate at the shock, $Q_i$.  If we assume that $Q_i$ evolves with time as $\propto \chi^{\beta^{\prime}}$ where $\beta^{\prime}$ is a constant, since the escaping CR spectrum per unit time is expressed as Eq.(\ref{escaping}), where $\phi(p)$ is proportional to $u_{\rm sh}f_{i,0}$, one can see that the time-integrated spectrum of escaping CRs would be 
\begin{eqnarray}
\int dt \frac{dN_{{\rm esc},i}}{dt}&\propto& \chi^{\beta^{\prime}+3}p_{i,m}^{\gamma-2} \nonumber \\
&\propto & p^{-(\gamma-2)-\frac{\beta^{\prime}+3}{\alpha}},
\end{eqnarray}
where we use $p_{i,m}\propto \chi^{-\alpha}$.  Considering that the power-law index of the on-site CR spectrum is $\sim \gamma-2$, we can see how $Q_i$ should increase with time as
\begin{eqnarray}
\beta^{\prime}=\beta-3.
\end{eqnarray}

\subsection{Observed CR Spectra from a Local Single SNR}

In this study, we try to explain the observed CR spectral hardening above $\sim 200~{\rm GV}$ by introducing a single local SNR surrounded by the ISM or a dense CSM.  The propagation of CR particles in the ISM can be described by the diffusion equation,
\begin{eqnarray}
\frac{\partial}{\partial t}f_{i,{\rm ISM}}(r,p,t)=D_{\rm ISM}(p)\nabla^2 f_{i,{\rm ISM}}(r,p,t)+q_i(r,p,t),
\end{eqnarray}
where $f_{i,{\rm ISM}}(r,p,t)$ is the distribution function of CR particles in the ISM at a distance $r$ from the source and the time $t$, with momentum $p$, and $q_i(r,p,t)$ is the CR injection rate from the source, which is located at $r=0$ (i.e., $q_i\propto \delta(r)$).  In our case, this diffusion equation can be solved as
\begin{eqnarray}
f_{i,{\rm ISM}}(r,p,t)=\int _0^t dt^{\prime} \frac{4\pi R_{\rm sh}(t^{\prime})^2 p^2 \left| \phi_i(p) \right|}{\left(4\pi D_{\rm ISM}t\right)^{3/2}}\exp \left( -\frac{r^2}{4D_{\rm ISM}t} \right). \label{observedsnr}
\end{eqnarray}.
The observed spectra can then be obtained by summing this distribution function and the background flux due to the myriad of sources in the Galaxy.

\section{Results and Discussion}
\subsection{Time integrated energy spectra of CR nuclei escaping the SNR}
\begin{figure}
 \begin{center}
  \includegraphics[width=80mm]{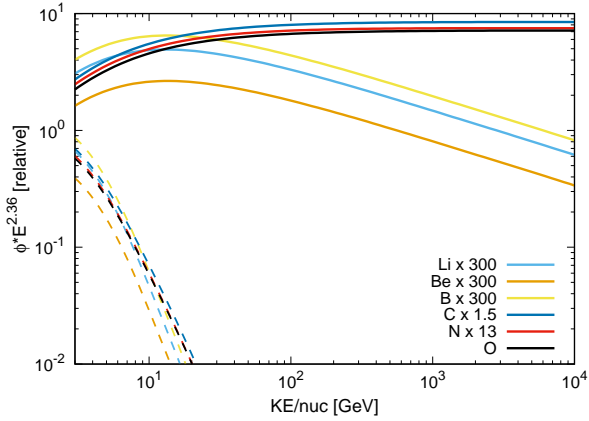}
 \end{center}
 \caption{Time-integrated energy spectra of CR nuclei escaping the SNR surrounded by a uniform interstellar medium ({\it solid lines}) and those once advected downstream and confined in the SNR ({\it dashed lines}).  The ambient density is assumed to be $0.1~{\rm cm}^{-1}$.}
 \label{fig1}
\end{figure}

\begin{figure}
 \begin{center}
  \includegraphics[width=80mm]{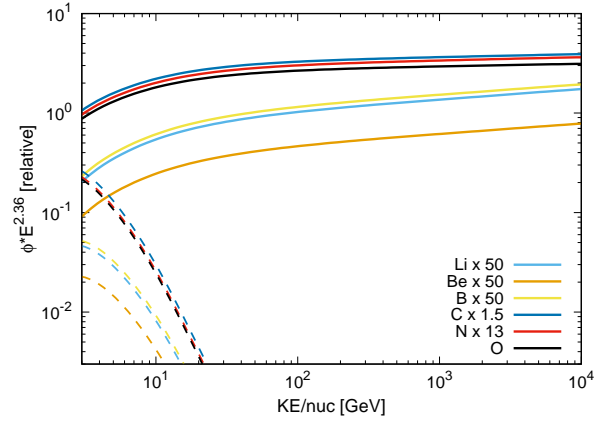}
 \end{center}
 \caption{Time-integrated energy spectra of CR nuclei escaping the SNR surrounded by a wind-like circumstellar medium ({\it solid lines}) and those once advected downstream and confined in the SNR ({\it dashed lines}).  The mass loss rate and wind velocity are assumed to be $3\times 10^{-3}M_{\odot}~{\rm yr}^{-1}$ and $100~{\rm km}~{\rm s}^{-1}$, respectively.}
 \label{fig2}
\end{figure}

Fig. 1 depicts the time-integrated spectra of CRs that have escaped the SNR surrounded by a uniform ISM during the Sedov phase and those that had been advected downstream and confined in the SNR until it was disrupted at the end of the Sedov phase\footnote{Here we assume that the advected CRs are released instantaneously when the SNR is disrupted, which may overestimate the effect of adiabatic cooling.}.  Here the ISM density is assumed as $0.1~{\rm cm}^{-3}$). Fig. 2 depicts the same spectra but in the case that the SNR is surrounded by a wind-like medium whose mass loss rate and velocity are $3\times 10^{-3}M_{\odot}~{\rm yr}^{-1}$ and $100~{\rm km}~{\rm s}^{-1}$, respectively.  In both cases, the explosion energy and ejecta mass of the supernova are assumed to be $10^{51}~{\rm erg}$ and 3 $M_{\odot}$ respectively, and the elemental abundance in CRs are assumed to be identical to that of the background CRs.  First of all, we can see that in both cases the contributions of CRs that had once been advected into the SNR are much lower than those escaping the SNR during the Sedov phase.  This means that we can consider only the CR particles that escape the SNR during the Sedov phase to discuss the observed spectral feature of CRs.  Second, in the former case, the spectra of escaping secondary nuclei (lithium, beryllium, and boron) are softer than those of the primary nuclei (carbon and oxygen), while in the latter case the spectra of secondary nuclei are a bit harder than those of primary nuclei.  These spectral features and the difference between two cases can be interpreted in the following way.  Since the secondary CR nuclei escaping the SNR are produced by the spallation of primary CR nuclei being accelerated at the shock, the number of CR particles escaping the SNR per unit time is proportional to $u_- f_{\rm pr}n_{\rm amb}t_{\rm int}$, where $f_{\rm pr}$ is the distribution function of their parent nuclei accelerated at the shock ($\chi^{\beta}$), $n_{\rm amb}$ is the density of the ambient gas (constant, or $\propto \chi^{-2}$), and $t_{\rm int}$ is the time during which their parent nuclei interact with the ambient gas before escape.  Especially, in the escape-limited regime, $t_{\rm int}$ is equal to their acceleration time in the SNR, $\sim D_i/u_-^2$.  Since we are assuming Bohm diffusion in the SNR, $D_i\propto p$, the intrinsic spectral index of secondary nuclei inside the SNR would be $s-1$.  On the other hand, the normalization factor of the secondary CR flux evolves with time as $\propto D_i/u_-^2\cdot \chi^{\beta}$ in the case of a uniform ISM and $\propto D_i/u_-^2 \cdot \chi^{-2} \cdot \chi^{\beta}$ in the case of a wind-like ambient medium.  Taking into account the time-dependence of the shell expansion velocity ($u_-\propto \chi^{-3/2}$ in the case of uniform ISM and $u_-\propto \chi^{-1/2}$ in the case with the wind-like ambient medium) and the diffusion coefficient (see the previous section), we can evaluate the spectral index of the distribution function of secondary CR nuclei escaping the SNR per unit time as 
\begin{eqnarray}
(s-1)+\frac{-3/2+\beta+(\alpha+5/2)}{\alpha}=s+\frac{\beta+1}{\alpha},
\end{eqnarray}
in the case of uniform ISM, and
\begin{eqnarray}
(s-1)+\frac{-1/2+\beta-2+(\alpha+3/2)}{\alpha}=s+\frac{\beta-1}{\alpha},
\end{eqnarray}
in the case of wind-like ambient medium.  Now we can see that in the former case the spectrum of secondary nuclei is softer, while in the latter case it is harder compared to that of the primary nuclei, regardless of the value of $\alpha$ or $\beta$.  This difference is caused mainly by the difference in the density profile of the ambient medium.  Generally, the CR particles with higher energy would escape the SNR earlier, so the time for primary CRs with higher energy to produce secondary CRs due to the interaction with ambient matter would be shorter.  This makes the number of escaping secondary CRs with higher energy smaller when the ambient matter distribution is uniform.  By contrast, when the SNR is surrounded by a wind-like CSM, the primary CRs escaping into the ISM earlier can interact with larger amount of matter than those escaping later, and then the number of secondary CRs produced by them is enhanced.  As a result, the energy spectrum of secondary CRs escaping the SNR would be harder than that of primary CRs.  Note that this spectral hardening of secondary nuclei is essentially different from what have been studied in \cite{2009PhRvL.103h1104M} and \cite{2014PhRvD..90f1301M}, in that they discussed the acceleration of secondary CRs produced inside the SNR assuming that the maximum energy of accelerated CRs is limited by the age of the SNR.  In this scenario, the primary CRs with higher energy can interact with ambient medium longer and produce more secondary CRs than those with lower energy. However, they did not take into account the energy-dependent escape of primary and secondary CRs, or the non-uniform distribution of ambient medium, both of which have been implied by recent observations of SNRs and SNe.

According to the recent measurements of secondary CR nuclei by AMS-02 \citep{PhysRevLett.120.021101}, the spectra of lithium, beryllium, and boron are hardened at $\sim 200~{\rm GV}$ as has been observed for primary CR nuclei, but they harden more than the primaries.  To account for such a feature, we will introduce the CR contribution from a past local SN. In this context, the case with a wind-like CSM, which can make the spectra of secondary CRs harder, is more relevant. In the following discussions, therefore, we will show results for the wind-like CSM case only. In addition to the elements shown in Figure 2, we also show the spectra of protons and helium nuclei expected from the SNR in the next section.

\subsection{Observed spectra of CR nuclei}
In our scenario, we assume that the flux of CR nuclei observed by AMS-02 is a superposition of two components: one is from a local SNR with a wind-like CSM that is supposed to reproduce the hardenings observed in the CR nuclei spectra at $\sim 200~{\rm GV}$, and the other is the background flux due to average SNRs without dense CSM (i.e., these SNRs do not produce hard secondary CRs efficiently inside themselves).  As for the first component, we can calculate its flux using Eq.(\ref{observedsnr}), choosing the parameters such as the distance and age of the SNR, total CR energy, CSM properties, abundance ratio in CRs, etc.  On the other hand, the background flux of $i$-th nuclei ${\mathcal N}_i$ is given recursively by
\begin{eqnarray}
{\mathcal N}_i(\varepsilon_k)=\frac{\sum_{i<j}\Gamma_{j\rightarrow i}{\mathcal N}_j(\varepsilon_k)+\mathcal{R}N_{{\rm esc},i}(\varepsilon_k)}{1/\tau_{{\rm esc},i}(\varepsilon_k)+\Gamma_i},
\end{eqnarray}
where ${\mathcal R}\simeq 0.03~{\rm yr}^{-1}$ is the Galactic SN rate and $\tau_{{\rm esc},i}(\varepsilon_k)$ is the timescale for an $i$-th nucleus with kinetic energy per nucleon of $\varepsilon_k$ to escape from the Galaxy, which is modeled using an expression of $D_{\rm ISM}$, 
\begin{eqnarray}
\tau_{{\rm esc},i}\simeq \frac{H^2}{D_{\rm ISM}(A_i\varepsilon_k/Z_i)},
\end{eqnarray}
where $H\simeq 3~{\rm kpc}$ is half the thickness of the Galactic halo.  In our calculation, we assume the diffusion coefficient in the ISM as $D_{\rm ISM}(R)=D_0(R/1~{\rm GV})^{0.45}$ where $D_0=2\times 10^{28}~{\rm cm}^2~{\rm s}^{-1}$ (see e.g., \citealp{2018ApJ...858...61B}).  

\begin{figure}
 \begin{center}
  \includegraphics[width=80mm]{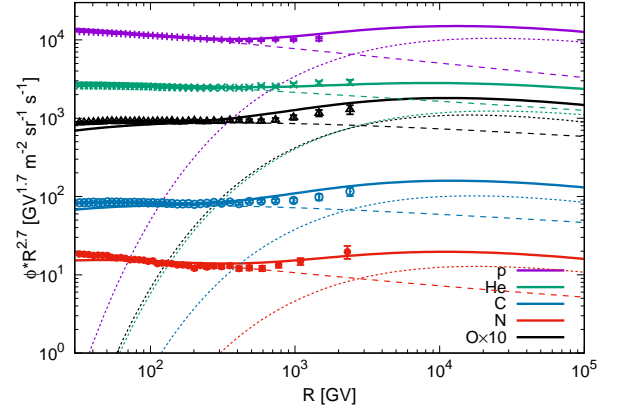}
 \end{center}
 \caption{Comparisons of our model spectra of CR oxygen, nitrogen, and carbon with AMS-02 data.  The background fluxes ({\it dashed lines}) and the contributions from our hypothetical past CSM-interacting SN ({\it dotted lines}) are also shown.}
 \label{fig3}
\end{figure}

\begin{figure}
 \begin{center}
  \includegraphics[width=80mm]{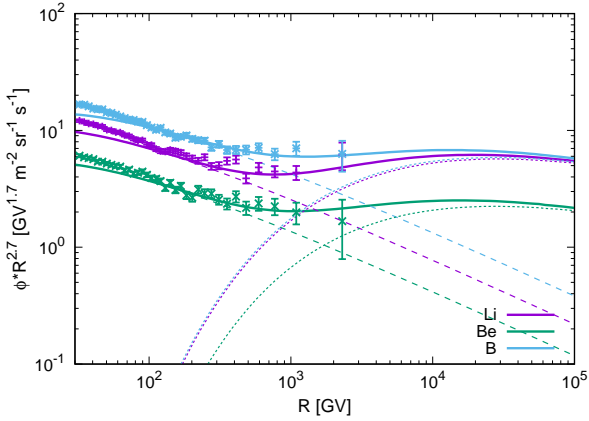}
 \end{center}
 \caption{Comparisons of our model spectra of CR boron, beryllium, and lithium with AMS-02 data.  The background fluxes ({\it dashed lines}) and the contributions from our hypothetical past CSM-interacting SN ({\it dotted lines}) are also shown.}
 \label{fig4}
\end{figure}

\begin{figure*}
  \begin{center}
    \begin{tabular}{cc}
        \begin{minipage}{0.33\hsize}
            \begin{center}
                \includegraphics[width=6cm]{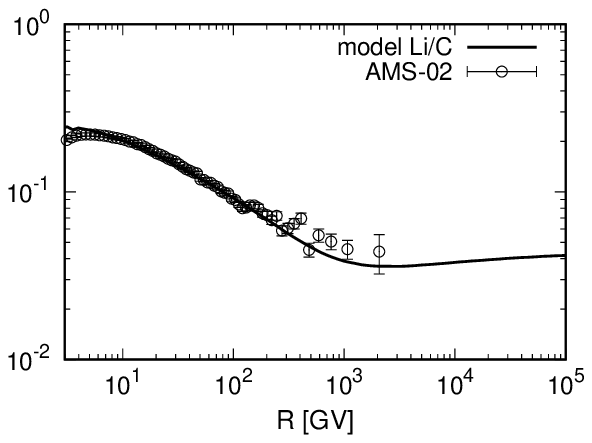}
                \hspace{1.6cm} 
            \end{center}
        \end{minipage}
        \begin{minipage}{0.33\hsize}
            \begin{center}
                \includegraphics[width=6cm]{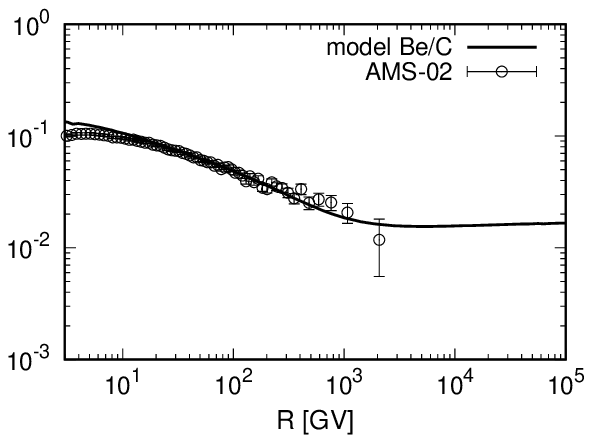}
                \hspace{1.6cm} 
            \end{center}
        \end{minipage}
        \begin{minipage}{0.33\hsize}
            \begin{center}
                \includegraphics[width=6cm]{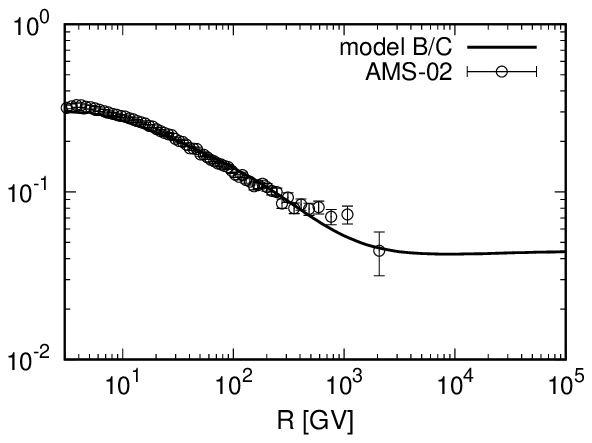}
                \hspace{1.6cm} 
            \end{center}
        \end{minipage}
    \end{tabular}
    \caption{Li/C ({\it left}), Be/C ({\it center}), and B/C ({\it right}) ratios in the observed CRs reported by AMS-02 along with our model predictions.}
    \label{fig5}
  \end{center}
\end{figure*}

\begin{table}
\begin{center}
  \begin{tabular}{l|c} \hline
    age & $1.6\times 10^5~{\rm year}$ \\ \hline 
    distance & $1.6~{\rm kpc}$  \\ \hline
    total energy & $10^{51}~{\rm erg}$ \\ \hline
    ejecta mass & $3~M_{\odot}$ \\ \hline
    mass loss rate & $2.5\times 10^{-3}M_{\odot}~{\rm yr}^{-1}$ \\ \hline
    wind velocity & $100~{\rm km}~{\rm s}^{-1}$ \\ \hline
  \end{tabular}
\end{center}
\caption{Properties of the past local SN we introduced to explain the AMS-02 data.}
\end{table}

Figure 3 depicts the observed spectra of proton, helium, carbon, nitrogen, and oxygen nuclei, and Figure 4 depicts the observed spectra of lithium, beryllium, and boron nuclei as functions of rigidity reported by AMS-02, fitted by our model.  Figure 5 depicts the predicted ratios of lithium to carbon, beryllium to carbon, and boron to carbon in the observed CRs as functions of rigidity compared with the ratios reported by AMS-02.  In making these plots, we assume a local SNR surrounded by dense wind-like CSM with parameters summarized in Table 1.  As for the elemental abundance in CRs, we assume that the fractions of carbon, nitrogen, and oxygen are three times as much as those in the background CRs.  We can see that the hard component appearing above $\sim 200~{\rm GV}$ in each spectrum is well explained by introducing this local SNR.  The important point here is that we can fit not only the spectra of primary CRs (i.e., oxygen and carbon) but also that of secondary CRs (i.e., lithium, beryllium and boron) by taking into account the production, acceleration, and escape of secondary CRs inside a local SNR surrounded by a dense CSM.  Another remarkable thing is that our model can be tested by the secondary to primary ratios at rigidity of $\gtrsim 1~{\rm TeV}$: within our model, they are predicted to show rising or flattening with rigidity.  Such a feature is never expected as long as we consider the ISM propagation effects as the origin of spectral hardening of CR nuclei.  As shown in Section 3.1, the energy spectra of secondary CR nuclei escaping a SNR surrounded by wind-like medium would be harder than those of the primary CR nuclei.  Since the observed CR spectra above $\sim 200~{\rm GV}$ are dominated by the contribution from a hypothetical SNR in our model, the predicted secondary-to-primary ratios in this rigidity range would reflect the hardening of secondary CRs escaping the SNR.  The secondary-to-primary ratios above $\sim~{\rm TeV}$ will be explored by future experiments such as AMS-100 \citep{2019NIMPA.94462561S}, which will provide a critical test for the existence of such a local SNR with dense wind-like CSM.

\subsection{Origin of the local SNR with a dense CSM}
While it is beyond the scope of this study to discuss in depth about the progenitor nature and detailed mass loss mechanism of the proposed local SNR, it is instructive to briefly account on its possible origin, and the feasibility of invoking such a local source without violating current observations. There exists a few possible scenarios for an enhanced mass loss rate during a certain period prior to core-collapse, such as giant eruptions at the envelope, envelope stripping in a binary system by Roche-lobe overflow (RLOF) to a companion star and a possible common envelope phase, and so on \citep[see, e.g.,][]{doi:10.1146/annurev-astro-081913-040025}. The hypothetical close-by SNR invoked here in particular can possibly be the result of a past SN of type Ib/c, originating from the explosion of a Wolf-Rayet (WR) star in a close binary system \citep[e.g.,][and reference therein]{2010ApJ...725..940Y, 2012MNRAS.424.2139D}. For example, a RLOF phase that lasts for $\sim 10^4$ yrs with a mass loss rate of $\sim 10^{-3}$ M$\mathrm{sun}$/yr and a velocity $\sim 100$ km/s can result in a dense CSM extending up to a radius $R_w \sim 1$ pc. Assuming an ejecta mass of $3 M_\odot$ typical for SN Ib/c, and for simplicity a $r^{-2}$ density profile for the CSM structure and the mass loss parameters in Table 1, the Sedov radius $R_S$ of the SNR is about $0.1$ pc.\footnote{We are ignoring any anisotropy and episodic history of the mass loss from the progenitor here.} This is much smaller than $R_w$, and the transition to Sedov phase happened at an age of roughly 10 yrs old. Therefore, the escape of the highest energy CRs must have occurred when the SNR blastwave was still in the midst of interacting with the dense CSM material, which is consistent with our picture above. 

On the other hand, the required chemical enhancement of heavy metal abundances in the CRs produced by this local accelerator is also consistent with a stripped envelope SN origin, in that the CSM in the vicinity of the ejecta of a type Ib/c progenitor is expected to be rich in metals in comparison to that around a SN IIP from the explosion of a red super-giant star \citep[e.g.,][]{yoon_2015}. The blast wave colliding with this metal-rich CSM shell in the early phase after SN should then be able to accelerate these heavy ions to high energies\footnote{We note that ejecta of SN Ib/c can reach trans-relativistic velocities right after explosions, for which particle acceleration at the shock can be modelled self-consistently using methods such as Monte-Carlo simulations \citep[e.g.,][]{Ellison_2013, 2015MNRAS.452..431W}. A more accurate approach with treatment of DSA at trans-relativistic shocks will be done in a future work.}. This interaction between the blastwave and a chemically-enriched CSM shell has been suggested by recent observations of type Ibn SNe \citep[e.g., see review in][]{2017hsn..book..403S}, or even the hypothetical type Icn that may be detected in the future. Meanwhile, at a current age of $1.6 \times 10^5$ yrs, the shock should have long died out already due to radiative loss, and the SNR should have merged with and become indistinguishable from the surrounding ISM, rendering it invisible in various wavebands at a distance $> 1$~kpc. A more detailed model for the evolution and broadband emission of such a stripped-envelope SNR interacting with a dense CSM using a method similar to \citet{2019ApJ...876...27Y} will be presented in a follow-up paper (Matsuoka et al., in preparation).

\subsection{About the nonlinear effects}
In this study, we assume that the CR particles in the SNR can be treated as test particles, and neglect the nonlinear effects from the back reaction of CRs on the incoming flow that is known to modify the shock structure and amplify the magnetic fields.  In general, we should take into account these nonlinear effects in the calculation of CR production by DSA coupled to the hydrodynamics, in a way similar to some existing frameworks in the literature \citep[e.g.,][]{LEN2012}.  First, the acceleration efficiency and maximum energies of CRs are expected to increase due to the nonlinear effects.  Second, shock modification by CR feedback can harden the CR spectra around the maximum energies to some extent (for reviews see \citealp{2001RPPh...64..429M, 2012SSRv..173..491S}). The inclusion of nonlinear effects in our model can therefore necessitate some adjustments of our parameters to maintain a good fit to the data. However, even in that case, our main conclusion that shock interaction of SNR with a dense wind in the CSM can explain the enhanced hardening of the secondary CR nuclei will not change.  In this work, we decide to focus on demonstrating this new possibility using a simpler test-particle picture, and  reserve a nonlinear description for a follow-up paper.

\section{Summary}
In this paper, we investigated the effect of the production of secondary CR nuclei at supernova remnants by the diffusive shock acceleration mechanism on the CR population measured near the Earth using an analytic approach. By including nuclear spallation effects and acceleration locally at the acceleration sites, we predicted the spectra of the accelerated ions escaping from the SNR shocks.  Our results show that the secondary CR nuclei escaping the SNR have a softer spectral index compared to the primary CRs when the SNR is surrounded by uniform ISM, while they have a harder spectral index when they are surrounded by the CSM with a wind-like density profile.  We show that, by introducing a past and relatively close-by SN event surrounded by a wind-like CSM, the current CR measurements including the spectral hardening recently discovered above a few $100$~GV in ion species up to oxygen can be successfully reproduced. We suggest that this past local accelerator can be a SN of type Ib/c with its progenitor enclosed by a CSM from pre-SN mass loss highly enriched in heavy metals. Our model also predicts a characteristic spectral flattening of CR secondary-to-primary ratios, such as Li/C, Be/C and so on, above a rigidity $\sim 1$~TV. Future CR measurements by next-generation experiments such as AMS-100 with an extended high energy range will put this to the test.       

\section*{Acknowledgement}
We are grateful to Kohta Murase for his comments.  N.K. acknowledges support by the Hakubi project at Kyoto University.  S.H.L. acknowledges support by JSPS Grant No. JP19K03913 and the World Premier International Research Center Initiative (WPI), MEXT, Japan. 
\clearpage
\bibliography{2ndary}{}

\begin{thebibliography}{}
\expandafter\ifx\csname natexlab\endcsname\relax\def\natexlab#1{#1}\fi
\providecommand{\url}[1]{\href{#1}{#1}}
\providecommand{\dodoi}[1]{doi:~\href{http://doi.org/#1}{\nolinkurl{#1}}}
\providecommand{\doeprint}[1]{\href{http://ascl.net/#1}{\nolinkurl{http://ascl.net/#1}}}
\providecommand{\doarXiv}[1]{\href{https://arxiv.org/abs/#1}{\nolinkurl{https://arxiv.org/abs/#1}}}

\bibitem[{{Adriani} {et~al.}(2009){Adriani}, {Barbarino}, {Bazilevskaya},
  {Bellotti}, {Boezio}, {Bogomolov}, {Bonechi}, {Bongi}, {Bonvicini}, {Bottai},
  {Bruno}, {Cafagna}, {Campana}, {Carlson}, {Casolino}, {Castellini}, {de
  Pascale}, {de Rosa}, {de Simone}, {di Felice}, {Galper}, {Grishantseva},
  {Hofverberg}, {Koldashov}, {Krutkov}, {Kvashnin}, {Leonov}, {Malvezzi},
  {Marcelli}, {Menn}, {Mikhailov}, {Mocchiutti}, {Orsi}, {Osteria}, {Papini},
  {Pearce}, {Picozza}, {Ricci}, {Ricciarini}, {Simon}, {Sparvoli},
  {Spillantini}, {Stozhkov}, {Vacchi}, {Vannuccini}, {Vasilyev}, {Voronov},
  {Yurkin}, {Zampa}, {Zampa}, \& {Zverev}}]{2009Natur.458..607A}
{Adriani}, O., {Barbarino}, G.~C., {Bazilevskaya}, G.~A., {et~al.} 2009, \nat,
  458, 607, \dodoi{10.1038/nature07942}

\bibitem[{{Adriani} {et~al.}(2011){Adriani}, {Barbarino}, {Bazilevskaya},
  {Bellotti}, {Boezio}, {Bogomolov}, {Bonechi}, {Bongi}, {Bonvicini},
  {Borisov}, {Bottai}, {Bruno}, {Cafagna}, {Campana}, {Carbone}, {Carlson},
  {Casolino}, {Castellini}, {Consiglio}, {De Pascale}, {De Santis}, {De
  Simone}, {Di Felice}, {Galper}, {Gillard}, {Grishantseva}, {Jerse},
  {Karelin}, {Koldashov}, {Krutkov}, {Kvashnin}, {Leonov}, {Malakhov},
  {Malvezzi}, {Marcelli}, {Mayorov}, {Menn}, {Mikhailov}, {Mocchiutti},
  {Monaco}, {Mori}, {Nikonov}, {Osteria}, {Palma}, {Papini}, {Pearce},
  {Picozza}, {Pizzolotto}, {Ricci}, {Ricciarini}, {Rossetto}, {Sarkar},
  {Simon}, {Sparvoli}, {Spillantini}, {Stozhkov}, {Vacchi}, {Vannuccini},
  {Vasilyev}, {Voronov}, {Yurkin}, {Wu}, {Zampa}, {Zampa}, \&
  {Zverev}}]{2011Sci...332...69A}
---. 2011, Science, 332, 69, \dodoi{10.1126/science.1199172}

\bibitem[{{Aguilar} {et~al.}(2013){Aguilar}, {Alberti}, {Alpat}, {Alvino},
  {Ambrosi}, {Andeen}, {Anderhub}, {Arruda}, {Azzarello}, {Bachlechner}, \&
  et~al.}]{2013PhRvL.110n1102A}
{Aguilar}, M., {Alberti}, G., {Alpat}, B., {et~al.} 2013, Physical Review
  Letters, 110, 141102, \dodoi{10.1103/PhysRevLett.110.141102}

\bibitem[{{Aguilar} {et~al.}(2015{\natexlab{a}}){Aguilar}, {Aisa}, {Alpat},
  {Alvino}, {Ambrosi}, {Andeen}, {Arruda}, {Attig}, {Azzarello}, {Bachlechner},
  \& et~al.}]{2015PhRvL.114q1103A}
{Aguilar}, M., {Aisa}, D., {Alpat}, B., {et~al.} 2015{\natexlab{a}}, Physical
  Review Letters, 114, 171103, \dodoi{10.1103/PhysRevLett.114.171103}

\bibitem[{{Aguilar} {et~al.}(2015{\natexlab{b}}){Aguilar}, {Aisa}, {Alpat},
  {Alvino}, {Ambrosi}, {Andeen}, {Arruda}, {Attig}, {Azzarello}, {Bachlechner},
  \& et~al.}]{2015PhRvL.115u1101A}
---. 2015{\natexlab{b}}, Physical Review Letters, 115, 211101,
  \dodoi{10.1103/PhysRevLett.115.211101}

\bibitem[{{Aguilar} {et~al.}(2016{\natexlab{a}}){Aguilar}, {Ali Cavasonza},
  {Ambrosi}, {Arruda}, {Attig}, {Aupetit}, {Azzarello}, {Bachlechner}, {Barao},
  {Barrau}, \& et~al.}]{2016PhRvL.117w1102A}
{Aguilar}, M., {Ali Cavasonza}, L., {Ambrosi}, G., {et~al.} 2016{\natexlab{a}},
  Physical Review Letters, 117, 231102, \dodoi{10.1103/PhysRevLett.117.231102}

\bibitem[{{Aguilar} {et~al.}(2016{\natexlab{b}}){Aguilar}, {Ali Cavasonza},
  {Alpat}, {Ambrosi}, {Arruda}, {Attig}, {Aupetit}, {Azzarello}, {Bachlechner},
  {Barao}, \& et~al.}]{2016PhRvL.117i1103A}
{Aguilar}, M., {Ali Cavasonza}, L., {Alpat}, B., {et~al.} 2016{\natexlab{b}},
  Physical Review Letters, 117, 091103, \dodoi{10.1103/PhysRevLett.117.091103}

\bibitem[{Aguilar {et~al.}(2017)Aguilar, Ali~Cavasonza, Alpat, Ambrosi, Arruda,
  Attig, Aupetit, Azzarello, Bachlechner, Barao, Barrau, Barrin, Bartoloni,
  Basara, Ba\ifmmode \mbox{\c{s}}\else \c{s}\fi{}e\ifmmode \breve{g}\else
  \u{g}\fi{}mez-du Pree, Battarbee, Battiston, Becker, Behlmann, Beischer,
  Berdugo, Bertucci, Bindel, Bindi, de~Boer, Bollweg, Bonnivard, Borgia,
  Boschini, Bourquin, Bueno, Burger, Burger, Cadoux, Cai, Capell, Caroff,
  Casaus, Castellini, Cervelli, Chae, Chang, Chen, Chen, Chen, Cheng, Chou,
  Choumilov, Choutko, Chung, Clark, Clavero, Coignet, Consolandi, Contin,
  Corti, Creus, Crispoltoni, Cui, Dadzie, Dai, Datta, Delgado, Della~Torre,
  Demakov, Demirk\"oz, Derome, Di~Falco, Dimiccoli, D\'{\i}az, von Doetinchem,
  Dong, Donnini, Duranti, D'Urso, Egorov, Eline, Eronen, Feng, Fiandrini,
  Fisher, Formato, Galaktionov, Gallucci, Garc\'{\i}a-L\'opez, Gargiulo, Gast,
  Gebauer, Gervasi, Ghelfi, Giovacchini, G\'omez-Coral, Gong, Goy, Grabski,
  Grandi, Graziani, Guo, Haino, Han, He, Heil, Hoffman, Hsieh, Huang, Huang,
  Huh, Incagli, Ionica, Jang, Jia, Jinchi, Kang, Kanishev, Khiali, Kim, Kim,
  Kirn, Konak, Kounina, Kounine, Koutsenko, Kulemzin, La~Vacca, Laudi,
  Laurenti, Lazzizzera, Lebedev, Lee, Lee, Leluc, Li, Li, Li, Li, Li, Li, Li,
  Lim, Lin, Lipari, Lippert, Liu, Liu, Lordello, Lu, Lu, Luebelsmeyer, Luo,
  Luo, Lyu, Machate, Ma\~n\'a, Mar\'{\i}n, Martin, Mart\'{\i}nez, Masi, Maurin,
  Menchaca-Rocha, Meng, Mikuni, Mo, Mott, Nelson, Ni, Nikonov, Nozzoli, Oliva,
  Orcinha, Palmonari, Palomares, Paniccia, Pauluzzi, Pensotti, Perrina, Phan,
  Picot-Clemente, Pilo, Pizzolotto, Plyaskin, Pohl, Poireau, Quadrani, Qi, Qin,
  Qu, R\"aih\"a, Rancoita, Rapin, Ricol, Rosier-Lees, Rozhkov, Rozza, Sagdeev,
  Schael, Schmidt, Schulz~von Dratzig, Schwering, Seo, Shan, Shi, Siedenburg,
  Son, Song, Tacconi, Tang, Tang, Tescaro, Ting, Ting, Tomassetti, Torsti,
  T\"urko\ifmmode~\breve{g}\else \u{g}\fi{}lu, Urban, Vagelli, Valente,
  Valtonen, V\'azquez~Acosta, Vecchi, Velasco, Vialle, Vitale, Vitillo, Wang,
  Wang, Wang, Wang, Wang, Wang, Wei, Weng, Whitman, Wu, Wu, Xiong, Xu, Yan,
  Yang, Yang, Yang, Yi, Yu, Yu, Zannoni, Zeissler, Zhang, Zhang, Zhang, Zhang,
  Zhang, Zhang, Zheng, Zhuang, Zhukov, Zichichi, Zimmermann, \&
  Zuccon}]{PhysRevLett.119.251101}
Aguilar, M., Ali~Cavasonza, L., Alpat, B., {et~al.} 2017, Phys. Rev. Lett.,
  119, 251101, \dodoi{10.1103/PhysRevLett.119.251101}

\bibitem[{Aguilar {et~al.}(2018)Aguilar, Ali~Cavasonza, Ambrosi, Arruda, Attig,
  Aupetit, Azzarello, Bachlechner, Barao, Barrau, Barrin, Bartoloni, Basara,
  Ba\ifmmode \mbox{\c{s}}\else \c{s}\fi{}e\ifmmode \breve{g}\else
  \u{g}\fi{}mez-du Pree, Battarbee, Battiston, Becker, Behlmann, Beischer,
  Berdugo, Bertucci, Bindel, Bindi, de~Boer, Bollweg, Bonnivard, Borgia,
  Boschini, Bourquin, Bueno, Burger, Burger, Cadoux, Cai, Capell, Caroff,
  Casaus, Castellini, Cervelli, Chae, Chang, Chen, Chen, Chen, Cheng, Chou,
  Choumilov, Choutko, Chung, Clark, Clavero, Coignet, Consolandi, Contin,
  Corti, Creus, Crispoltoni, Cui, Dadzie, Dai, Datta, Delgado, Della~Torre,
  Demirk\"oz, Derome, Di~Falco, Dimiccoli, D\'{\i}az, von Doetinchem, Dong,
  Donnini, Duranti, D'Urso, Egorov, Eline, Eronen, Feng, Fiandrini, Fisher,
  Formato, Galaktionov, Gallucci, Garc\'{\i}a-L\'opez, Gargiulo, Gast, Gebauer,
  Gervasi, Ghelfi, Giovacchini, G\'omez-Coral, Gong, Goy, Grabski, Grandi,
  Graziani, Guo, Haino, Han, He, Heil, Hsieh, Huang, Huang, Huh, Incagli,
  Ionica, Jang, Jia, Jinchi, Kang, Kanishev, Khiali, Kim, Kim, Kirn, Konak,
  Kounina, Kounine, Koutsenko, Kulemzin, La~Vacca, Laudi, Laurenti, Lazzizzera,
  Lebedev, Lee, Lee, Leluc, Li, Li, Li, Li, Li, Li, Li, Lim, Lin, Lipari,
  Lippert, Liu, Liu, Lordello, Lu, Lu, Luebelsmeyer, Luo, Luo, Lyu, Machate,
  Ma\~n\'a, Mar\'{\i}n, Martin, Mart\'{\i}nez, Masi, Maurin, Menchaca-Rocha,
  Meng, Mikuni, Mo, Mott, Nelson, Ni, Nikonov, Nozzoli, Oliva, Orcinha,
  Palermo, Palmonari, Palomares, Paniccia, Pauluzzi, Pensotti, Perrina, Phan,
  Picot-Clemente, Pilo, Pizzolotto, Plyaskin, Pohl, Poireau, Quadrani, Qi, Qin,
  Qu, R\"aih\"a, Rancoita, Rapin, Ricol, Rosier-Lees, Rozhkov, Rozza, Sagdeev,
  Schael, Schmidt, Schulz~von Dratzig, Schwering, Seo, Shan, Shi, Siedenburg,
  Son, Song, Tacconi, Tang, Tang, Tescaro, Ting, Ting, Tomassetti, Torsti,
  T\"urko\ifmmode~\breve{g}\else \u{g}\fi{}lu, Urban, Vagelli, Valente,
  Valtonen, V\'azquez~Acosta, Vecchi, Velasco, Vialle, Vitale, Wang, Wang,
  Wang, Wang, Wang, Wang, Wei, Weng, Whitman, Wu, Wu, Xiong, Xu, Yan, Yang,
  Yang, Yang, Yi, Yu, Yu, Zannoni, Zeissler, Zhang, Zhang, Zhang, Zhang, Zhang,
  Zhang, Zheng, Zhuang, Zhukov, Zichichi, Zimmermann, \&
  Zuccon}]{PhysRevLett.120.021101}
Aguilar, M., Ali~Cavasonza, L., Ambrosi, G., {et~al.} 2018, Phys. Rev. Lett.,
  120, 021101, \dodoi{10.1103/PhysRevLett.120.021101}

\bibitem[{{Ahlers} {et~al.}(2009){Ahlers}, {Mertsch}, \&
  {Sarkar}}]{2009PhRvD..80l3017A}
{Ahlers}, M., {Mertsch}, P., \& {Sarkar}, S. 2009, \prd, 80, 123017,
  \dodoi{10.1103/PhysRevD.80.123017}

\bibitem[{{Ahn} {et~al.}(2010){Ahn}, {Allison}, {Bagliesi}, {Beatty},
  {Bigongiari}, {Childers}, {Conklin}, {Coutu}, {DuVernois}, {Ganel}, {Han},
  {Jeon}, {Kim}, {Lee}, {Lutz}, {Maestro}, {Malinin}, {Marrocchesi}, {Minnick},
  {Mognet}, {Nam}, {Nam}, {Nutter}, {Park}, {Park}, {Seo}, {Sina}, {Wu},
  {Yang}, {Yoon}, {Zei}, \& {Zinn}}]{2010ApJ...714L..89A}
{Ahn}, H.~S., {Allison}, P., {Bagliesi}, M.~G., {et~al.} 2010, \apjl, 714, L89,
  \dodoi{10.1088/2041-8205/714/1/L89}

\bibitem[{{Bell}(1978)}]{1978MNRAS.182..147B}
{Bell}, A.~R. 1978, \mnras, 182, 147, \dodoi{10.1093/mnras/182.2.147}

\bibitem[{{Bell} {et~al.}(2013){Bell}, {Schure}, {Reville}, \&
  {Giacinti}}]{2013MNRAS.431..415B}
{Bell}, A.~R., {Schure}, K.~M., {Reville}, B., \& {Giacinti}, G. 2013, \mnras,
  431, 415, \dodoi{10.1093/mnras/stt179}

\bibitem[{{Berezinskii} {et~al.}(1990){Berezinskii}, {Bulanov}, {Dogiel}, \&
  {Ptuskin}}]{1990acr..book.....B}
{Berezinskii}, V.~S., {Bulanov}, S.~V., {Dogiel}, V.~A., \& {Ptuskin}, V.~S.
  1990, {Astrophysics of cosmic rays}

\bibitem[{{Blandford} \& {Eichler}(1987)}]{1987PhR...154....1B}
{Blandford}, R., \& {Eichler}, D. 1987, \physrep, 154, 1,
  \dodoi{10.1016/0370-1573(87)90134-7}

\bibitem[{{Boschini} {et~al.}(2018){Boschini}, {Della Torre}, {Gervasi},
  {Grandi}, {J{\'o}hannesson}, {La Vacca}, {Masi}, {Moskalenko}, {Pensotti},
  {Porter}, {Quadrani}, {Rancoita}, {Rozza}, \&
  {Tacconi}}]{2018ApJ...858...61B}
{Boschini}, M.~J., {Della Torre}, S., {Gervasi}, M., {et~al.} 2018, \apj, 858,
  61, \dodoi{10.3847/1538-4357/aabc54}

\bibitem[{{Boschini} {et~al.}(2020){Boschini}, {Della Torre}, {Gervasi},
  {Grandi}, {J{\o}hannesson}, {La Vacca}, {Masi}, {Moskalenko}, {Pensotti},
  {Porter}, {Quadrani}, {Rancoita}, {Rozza}, \&
  {Tacconi}}]{2020ApJ...889..167B}
---. 2020, \apj, 889, 167, \dodoi{10.3847/1538-4357/ab64f1}

\bibitem[{{Bresci} {et~al.}(2019){Bresci}, {Amato}, {Blasi}, \&
  {Morlino}}]{2019MNRAS.488.2068B}
{Bresci}, V., {Amato}, E., {Blasi}, P., \& {Morlino}, G. 2019, \mnras, 488,
  2068, \dodoi{10.1093/mnras/stz1806}

\bibitem[{{Caprioli} {et~al.}(2009){Caprioli}, {Blasi}, \&
  {Amato}}]{2009MNRAS.396.2065C}
{Caprioli}, D., {Blasi}, P., \& {Amato}, E. 2009, \mnras, 396, 2065,
  \dodoi{10.1111/j.1365-2966.2008.14298.x}

\bibitem[{{Caprioli} {et~al.}(2010){Caprioli}, {Kang}, {Vladimirov}, \&
  {Jones}}]{2010MNRAS.407.1773C}
{Caprioli}, D., {Kang}, H., {Vladimirov}, A.~E., \& {Jones}, T.~W. 2010,
  \mnras, 407, 1773, \dodoi{10.1111/j.1365-2966.2010.17013.x}

\bibitem[{{Cholis} \& {Hooper}(2014)}]{2014PhRvD..89d3013C}
{Cholis}, I., \& {Hooper}, D. 2014, \prd, 89, 043013,
  \dodoi{10.1103/PhysRevD.89.043013}

\bibitem[{{Cowsik} \& {Burch}(2010)}]{2010PhRvD..82b3009C}
{Cowsik}, R., \& {Burch}, B. 2010, \prd, 82, 023009,
  \dodoi{10.1103/PhysRevD.82.023009}

\bibitem[{{Cowsik} {et~al.}(2014){Cowsik}, {Burch}, \&
  {Madziwa-Nussinov}}]{2014ApJ...786..124C}
{Cowsik}, R., {Burch}, B., \& {Madziwa-Nussinov}, T. 2014, \apj, 786, 124,
  \dodoi{10.1088/0004-637X/786/2/124}

\bibitem[{{Cowsik} \& {Madziwa-Nussinov}(2016)}]{2016ApJ...827..119C}
{Cowsik}, R., \& {Madziwa-Nussinov}, T. 2016, \apj, 827, 119,
  \dodoi{10.3847/0004-637X/827/2/119}

\bibitem[{{Cristofari} {et~al.}(2020){Cristofari}, {Renaud}, {Marcowith},
  {Dwarkadas}, \& {Tatischeff}}]{2020MNRAS.494.2760C}
{Cristofari}, P., {Renaud}, M., {Marcowith}, A., {Dwarkadas}, V.~V., \&
  {Tatischeff}, V. 2020, \mnras, 494, 2760, \dodoi{10.1093/mnras/staa984}

\bibitem[{{Dessart} {et~al.}(2012){Dessart}, {Hillier}, {Li}, \&
  {Woosley}}]{2012MNRAS.424.2139D}
{Dessart}, L., {Hillier}, D.~J., {Li}, C., \& {Woosley}, S. 2012, \mnras, 424,
  2139, \dodoi{10.1111/j.1365-2966.2012.21374.x}

\bibitem[{{Drury}(2011)}]{2011MNRAS.415.1807D}
{Drury}, L.~O. 2011, \mnras, 415, 1807,
  \dodoi{10.1111/j.1365-2966.2011.18824.x}

\bibitem[{Ellison {et~al.}(2013)Ellison, Warren, \& Bykov}]{Ellison_2013}
Ellison, D.~C., Warren, D.~C., \& Bykov, A.~M. 2013, The Astrophysical Journal,
  776, 46, \dodoi{10.1088/0004-637x/776/1/46}

\bibitem[{{Gabici} {et~al.}(2009){Gabici}, {Aharonian}, \&
  {Casanova}}]{2009MNRAS.396.1629G}
{Gabici}, S., {Aharonian}, F.~A., \& {Casanova}, S. 2009, \mnras, 396, 1629,
  \dodoi{10.1111/j.1365-2966.2009.14832.x}

\bibitem[{{Kachelrie{\ss}} \& {Ostapchenko}(2013)}]{2013PhRvD..87d7301K}
{Kachelrie{\ss}}, M., \& {Ostapchenko}, S. 2013, \prd, 87, 047301,
  \dodoi{10.1103/PhysRevD.87.047301}

\bibitem[{{Kawanaka} {et~al.}(2011){Kawanaka}, {Ioka}, {Ohira}, \&
  {Kashiyama}}]{2011ApJ...729...93K}
{Kawanaka}, N., {Ioka}, K., {Ohira}, Y., \& {Kashiyama}, K. 2011, \apj, 729,
  93, \dodoi{10.1088/0004-637X/729/2/93}

\bibitem[{{Kawanaka} \& {Yanagita}(2018)}]{2018PhRvL.120d1103K}
{Kawanaka}, N., \& {Yanagita}, S. 2018, \prl, 120, 041103,
  \dodoi{10.1103/PhysRevLett.120.041103}

\bibitem[{{Kolmogorov}(1941)}]{1941DoSSR..30..301K}
{Kolmogorov}, A. 1941, Akademiia Nauk SSSR Doklady, 30, 301

\bibitem[{{Lee} {et~al.}(2012){Lee}, {Ellison}, \& {Nagataki}}]{LEN2012}
{Lee}, S.-H., {Ellison}, D.~C., \& {Nagataki}, S. 2012, \apj, 750, 156.
\newblock \doarXiv{1203.3614}

\bibitem[{{Lucek} \& {Bell}(2000)}]{2000MNRAS.314...65L}
{Lucek}, S.~G., \& {Bell}, A.~R. 2000, \mnras, 314, 65,
  \dodoi{10.1046/j.1365-8711.2000.03363.x}

\bibitem[{{Malkov} \& {Drury}(2001)}]{2001RPPh...64..429M}
{Malkov}, M.~A., \& {Drury}, L.~O. 2001, Reports on Progress in Physics, 64,
  429, \dodoi{10.1088/0034-4885/64/4/201}

\bibitem[{{Malkov} \& {Moskalenko}(2021)}]{2021ApJ...911..151M}
{Malkov}, M.~A., \& {Moskalenko}, I.~V. 2021, \apj, 911, 151,
  \dodoi{10.3847/1538-4357/abe855}

\bibitem[{Matsuoka \& Maeda(2020)}]{Matsuoka_2020}
Matsuoka, T., \& Maeda, K. 2020, \apj, 898, 158,
  \dodoi{10.3847/1538-4357/ab9c1b}

\bibitem[{{Matsuoka} {et~al.}(2019){Matsuoka}, {Maeda}, {Lee}, \&
  {Yasuda}}]{2019ApJ...885...41M}
{Matsuoka}, T., {Maeda}, K., {Lee}, S.-H., \& {Yasuda}, H. 2019, \apj, 885, 41,
  \dodoi{10.3847/1538-4357/ab4421}

\bibitem[{{Mertsch} \& {Sarkar}(2009)}]{2009PhRvL.103h1104M}
{Mertsch}, P., \& {Sarkar}, S. 2009, Physical Review Letters, 103, 081104,
  \dodoi{10.1103/PhysRevLett.103.081104}

\bibitem[{{Mertsch} \& {Sarkar}(2014)}]{2014PhRvD..90f1301M}
---. 2014, \prd, 90, 061301, \dodoi{10.1103/PhysRevD.90.061301}

\bibitem[{{Mertsch} {et~al.}(2020){Mertsch}, {Vittino}, \&
  {Sarkar}}]{2020arXiv201212853M}
{Mertsch}, P., {Vittino}, A., \& {Sarkar}, S. 2020, arXiv e-prints,
  arXiv:2012.12853.
\newblock \doarXiv{2012.12853}

\bibitem[{{Moriya} {et~al.}(2014){Moriya}, {Maeda}, {Taddia}, {Sollerman},
  {Blinnikov}, \& {Sorokina}}]{2014MNRAS.439.2917M}
{Moriya}, T.~J., {Maeda}, K., {Taddia}, F., {et~al.} 2014, \mnras, 439, 2917,
  \dodoi{10.1093/mnras/stu163}

\bibitem[{{Murase} {et~al.}(2019){Murase}, {Franckowiak}, {Maeda}, {Margutti},
  \& {Beacom}}]{2019ApJ...874...80M}
{Murase}, K., {Franckowiak}, A., {Maeda}, K., {Margutti}, R., \& {Beacom},
  J.~F. 2019, \apj, 874, 80, \dodoi{10.3847/1538-4357/ab0422}

\bibitem[{{Murase} {et~al.}(2011){Murase}, {Thompson}, {Lacki}, \&
  {Beacom}}]{2011PhRvD..84d3003M}
{Murase}, K., {Thompson}, T.~A., {Lacki}, B.~C., \& {Beacom}, J.~F. 2011, \prd,
  84, 043003, \dodoi{10.1103/PhysRevD.84.043003}

\bibitem[{{Murase} {et~al.}(2014){Murase}, {Thompson}, \&
  {Ofek}}]{2014MNRAS.440.2528M}
{Murase}, K., {Thompson}, T.~A., \& {Ofek}, E.~O. 2014, \mnras, 440, 2528,
  \dodoi{10.1093/mnras/stu384}

\bibitem[{{Niu}(2021)}]{2021ChPhC..45d1004N}
{Niu}, J.-S. 2021, Chinese Physics C, 45, 041004,
  \dodoi{10.1088/1674-1137/abe03d}

\bibitem[{{Ohira} \& {Ioka}(2011)}]{2011ApJ...729L..13O}
{Ohira}, Y., \& {Ioka}, K. 2011, \apjl, 729, L13,
  \dodoi{10.1088/2041-8205/729/1/L13}

\bibitem[{{Ohira} {et~al.}(2016){Ohira}, {Kawanaka}, \&
  {Ioka}}]{2016PhRvD..93h3001O}
{Ohira}, Y., {Kawanaka}, N., \& {Ioka}, K. 2016, \prd, 93, 083001,
  \dodoi{10.1103/PhysRevD.93.083001}

\bibitem[{{Ohira} {et~al.}(2010){Ohira}, {Murase}, \&
  {Yamazaki}}]{2010A&A...513A..17O}
{Ohira}, Y., {Murase}, K., \& {Yamazaki}, R. 2010, \aap, 513, A17,
  \dodoi{10.1051/0004-6361/200913495}

\bibitem[{{Ohira} {et~al.}(2011){Ohira}, {Murase}, \&
  {Yamazaki}}]{2011MNRAS.410.1577O}
---. 2011, \mnras, 410, 1577, \dodoi{10.1111/j.1365-2966.2010.17539.x}

\bibitem[{{Ohira} {et~al.}(2012){Ohira}, {Yamazaki}, {Kawanaka}, \&
  {Ioka}}]{2012MNRAS.427...91O}
{Ohira}, Y., {Yamazaki}, R., {Kawanaka}, N., \& {Ioka}, K. 2012, \mnras, 427,
  91, \dodoi{10.1111/j.1365-2966.2012.21908.x}

\bibitem[{{Ptuskin} \& {Zirakashvili}(2003)}]{2003A&A...403....1P}
{Ptuskin}, V.~S., \& {Zirakashvili}, V.~N. 2003, \aap, 403, 1,
  \dodoi{10.1051/0004-6361:20030323}

\bibitem[{{Ptuskin} \& {Zirakashvili}(2005)}]{2005A&A...429..755P}
---. 2005, \aap, 429, 755, \dodoi{10.1051/0004-6361:20041517}

\bibitem[{{Reville} {et~al.}(2009){Reville}, {Kirk}, \&
  {Duffy}}]{2009ApJ...694..951R}
{Reville}, B., {Kirk}, J.~G., \& {Duffy}, P. 2009, \apj, 694, 951,
  \dodoi{10.1088/0004-637X/694/2/951}

\bibitem[{{Schael} {et~al.}(2019){Schael}, {Atanasyan}, {Berdugo}, {Bretz},
  {Czupalla}, {Dachwald}, {von Doetinchem}, {Duranti}, {Gast}, {Karpinski},
  {Kirn}, {L{\"u}belsmeyer}, {Ma{\~n}a}, {Marrocchesi}, {Mertsch},
  {Moskalenko}, {Schervan}, {Schluse}, {Schr{\"o}der}, {Schultz von Dratzig},
  {Senatore}, {Spies}, {Wakely}, {Wlochal}, {Uglietti}, \&
  {Zimmermann}}]{2019NIMPA.94462561S}
{Schael}, S., {Atanasyan}, A., {Berdugo}, J., {et~al.} 2019, Nuclear
  Instruments and Methods in Physics Research A, 944, 162561,
  \dodoi{10.1016/j.nima.2019.162561}

\bibitem[{{Schure} {et~al.}(2012){Schure}, {Bell}, {O'C Drury}, \&
  {Bykov}}]{2012SSRv..173..491S}
{Schure}, K.~M., {Bell}, A.~R., {O'C Drury}, L., \& {Bykov}, A.~M. 2012, \ssr,
  173, 491, \dodoi{10.1007/s11214-012-9871-7}

\bibitem[{Smith(2014)}]{doi:10.1146/annurev-astro-081913-040025}
Smith, N. 2014, Annual Review of Astronomy and Astrophysics, 52, 487,
  \dodoi{10.1146/annurev-astro-081913-040025}

\bibitem[{{Smith}(2017)}]{2017hsn..book..403S}
{Smith}, N. 2017, {Interacting Supernovae: Types IIn and Ibn}, ed. A.~W.
  {Alsabti} \& P.~{Murdin}, 403, \dodoi{10.1007/978-3-319-21846-5_38}

\bibitem[{{Tomassetti}(2012)}]{2012ApJ...752L..13T}
{Tomassetti}, N. 2012, \apjl, 752, L13, \dodoi{10.1088/2041-8205/752/1/L13}

\bibitem[{{Tomassetti}(2015)}]{2015PhRvD..92h1301T}
---. 2015, \prd, 92, 081301, \dodoi{10.1103/PhysRevD.92.081301}

\bibitem[{{Wang} {et~al.}(2019){Wang}, {Huang}, \& {Li}}]{2019ApJ...872..157W}
{Wang}, K., {Huang}, T.-Q., \& {Li}, Z. 2019, \apj, 872, 157,
  \dodoi{10.3847/1538-4357/aaffd9}

\bibitem[{{Warren} {et~al.}(2015){Warren}, {Ellison}, {Bykov}, \&
  {Lee}}]{2015MNRAS.452..431W}
{Warren}, D.~C., {Ellison}, D.~C., {Bykov}, A.~M., \& {Lee}, S.-H. 2015,
  \mnras, 452, 431, \dodoi{10.1093/mnras/stv1304}

\bibitem[{{Yan} {et~al.}(2012){Yan}, {Lazarian}, \&
  {Schlickeiser}}]{2012ApJ...745..140Y}
{Yan}, H., {Lazarian}, A., \& {Schlickeiser}, R. 2012, \apj, 745, 140,
  \dodoi{10.1088/0004-637X/745/2/140}

\bibitem[{{Yasuda} \& {Lee}(2019)}]{2019ApJ...876...27Y}
{Yasuda}, H., \& {Lee}, S.-H. 2019, \apj, 876, 27,
  \dodoi{10.3847/1538-4357/ab13ab}

\bibitem[{Yoon(2015)}]{yoon_2015}
Yoon, S.-C. 2015, Publications of the Astronomical Society of Australia, 32,
  e015, \dodoi{10.1017/pasa.2015.16}

\bibitem[{{Yoon} {et~al.}(2010){Yoon}, {Woosley}, \&
  {Langer}}]{2010ApJ...725..940Y}
{Yoon}, S.~C., {Woosley}, S.~E., \& {Langer}, N. 2010, \apj, 725, 940,
  \dodoi{10.1088/0004-637X/725/1/940}

\bibitem[{{Yuan} {et~al.}(2020){Yuan}, {Zhu}, {Bi}, \&
  {Wei}}]{2020JCAP...11..027Y}
{Yuan}, Q., {Zhu}, C.-R., {Bi}, X.-J., \& {Wei}, D.-M. 2020, \jcap, 2020, 027,
  \dodoi{10.1088/1475-7516/2020/11/027}

\bibitem[{{Zirakashvili} \& {Ptuskin}(2008)}]{2008ApJ...678..939Z}
{Zirakashvili}, V.~N., \& {Ptuskin}, V.~S. 2008, \apj, 678, 939,
  \dodoi{10.1086/529580}

\bibitem[{{Zirakashvili} \& {Ptuskin}(2016)}]{2016APh....78...28Z}
---. 2016, Astroparticle Physics, 78, 28,
  \dodoi{10.1016/j.astropartphys.2016.02.004}

\end{thebibliography}
\bibliographystyle{aasjournal} 

\end{document}